\documentstyle[12pt,aasms4]{article}

\begin{document}

\title{Gravitational Radiation From Globular Clusters}
\author{M. Benacquista}
\affil{Montana State University - Billings, Billings, MT 59101}
\authoremail{sci_quista@vixen.emcmt.edu}

\begin{abstract}
Space-based gravitational wave detectors will have the ability to observe continuous low frequency gravitational radiation from binary star systems. They can determine the direction to continuous sources with an angular resolution approaching tens of arcminutes. This resolution should be sufficient to identify binary sources as members of some nearby globular clusters. Thus, gravitational radiation can be used to determine the population of hard binaries in globular clusters. For particularly hard binaries, the orbital period may change as a result of gravitational wave emission. If one of these binaries can be identified with a globular cluster, then the distance to that cluster can be determined. Thus, gravitational radiation may provide reddening independent distance measurements to globular clusters and the RR-Lyrae stars that inhabit them.
\end{abstract}


\section{Introduction}
Space-based gravitational wave detectors will be sensitive to gravitational radiation in a low-frequency band of about $10^{-4}$ to 1 Hz. There are currently two proposed projects, LISA and OMEGA, which both share several common features. They consist of multiple space-craft arranged in an equilateral triangle to form the arms of laser interferometers, and the plane of the triangle is at an angle with respect to the plane of the ecliptic. Either detector will orbit the sun with a period of one year. LISA differs from OMEGA in that LISA's constellation of spacecraft is in a heliocentric orbit approximately $20\arcdeg$ behind the earth and the orientation of the plane varies during the course of the detector's orbit about the sun. The armlength of LISA is expected to be about $5 \times 10^6$ km. The spacecraft in OMEGA will be in a geocentric orbit with an armlength of $1 \times 10^6$ km. The detection of a passing gravitational wave is then achieved by measuring a change in the armlength of the interferometers. The common features allow either detector to have similar sensitivity to the distance and position of continuous binary sources of gravitational radiation.

The detected signal from a continuous source of gravitational radiation will have frequency and phase modulation which can be used to determined the distance and position of the source. In order to detect these sources, one must search through the data for a signal which matches a template incorporating all the expected properties of the source. The template is specific to a source with a given frequency, angular position, and distance. Consequently, one must sift through the data with a separate template for each possible source in the sky. Given the angular resolution expected for most reasonably strong sources, this may require looking at up to 1000 different patches in the sky for initial identification. Additional refinement will be necessary to locate the source to an area inversely proportional to the square of the signal to noise ratio (Schutz, 1997b)\markcite{schutz2}. It is expected that binary white dwarfs in the disk of the galaxy will produce a confusion limited background noise for frequencies below about 0.05 Hz (Postnov \& Prokhorov, 1998)\markcite{postnov}. Thus, it is possible that only the strongest sources in the galactic plane will be detectable above this background.

Globular clusters provide excellent targets for such searches since a number of them lie outside the galactic plane and they contain $10^5$ to $10^7$ stars. The angular size of most globular clusters is well within the angular resolution expected of either LISA or OMEGA. Thus, the initial search for sources can be bypassed and the additional refinement can take place immediately for sources in globular clusters. Furthermore, globular clusters should provide an excess of high-frequency, double degenerate binary systems which are strong sources of gravitational radiation. The age of globular clusters implies that there should be a number of evolved, compact objects such as neutron stars and white dwarfs. The dynamics of globular cluster evolution is expected to favor the creation of hard binaries. This possibility of finding sufficiently strong sources which can be identified with specific globular clusters also allows for the possibility of determining the distance to these clusters using gravitational radiation. If the distance to a source can be determined, gravitational radiation will provide a distance measurement which is independent of electromagnetic observation techniques.

\section{Signal Strength}
The strain amplitude due to a close binary system at a distance r is given by \markcite{schutz1}(Schutz 1997a):
\begin{equation}\label{strainamp}
h = 2 (4 \pi)^{1/3} \frac{G^{5/3}}{c^4} f_{gw}^{2/3} m_1 m_2 (m_1 + m_2)^{-1/3} \frac{1}{r},
\end{equation}
where the gravitational wave frequency, $f_{gw}$, is twice the orbital frequency of the binary. For an interferometer whose arms form an equilateral triangle, the measured strain amplitude will be reduced by a factor of $\sqrt{3}/2$ \markcite{cutler}(Cutler 1998). There are 16 globular clusters within about 5 kpc of the sun, and two very good candidates (47 Tuc and NGC 6752) which are both at about 4 kpc. At this distance, the measured signal for a variety of binary systems with orbital periods ranging from 2000 s to 20 s will have signal strengths well above the noise levels for either LISA or OMEGA. Table~\ref{ddbtype} lists the properties of six possible double degenerate binaries which may be found in a globular cluster and were used to determine the signal strengths.
\placetable{ddbtype}

In general, the binaries in Table~\ref{ddbtype} will have a lower limit to their orbital period. This critical period $(P_c)$ occurs at the onset of Roche lobe overflow. The critical period is calculated using the effective Roche Lobe radius (Eggleton 1983)\markcite{eggleton}:
\begin{equation}
r_L = \frac{0.49 q^{2/3}}{0.6 q^{2/3} + \ln{1 + q^{1/3}}}
\end{equation}
where q is the mass ratio $m_2/m_1$, and the following equations are used for the radii of white dwarfs and neutron stars:
\begin{equation}
R_{wd} = \frac{3\hbar^2}{5Gm_e}\left(\frac{9\pi}{4m_p^{5/2}}\right)^{2/3} M^{-1/3}
\end{equation}
\begin{equation}
R_{ns} = \frac{6\hbar^2}{5Gm_n}\left(\frac{9\pi}{4m_p^{5/2}}\right)^{2/3} M^{-1/3}
\end{equation}
where $M$ is the mass of the star and $m_e$, $m_n$, and $m_p$ are the masses of the electron, neutron, and proton. Equating the equivalent Roche lobe radius with the star radius gives the critical separation and thus the critical period. The periods calculated in this way are slightly larger than those given by \markcite{evans} Evans et al. (1987) for similar systems.

The signal strengths of the six types of binaries listed in Table~\ref{ddbtype} are shown in Figure~\ref{snrplot} for a globular cluster at 4 kpc from the sun. Since these systems will not change their frequencies significantly during the observation period, the signal to noise ratio can be comfortably estimated by taking the ratio of the signal strength to the noise level at a single frequency. Any binary systems containing a neutron star and a second degenerate object with an orbital period below 650 s will have a signal to noise ratio above 100. Only He-He systems will not be seen with signal to noise ratios above 100. Thus, if close double degenerate binaries exist in globular clusters, they will be easily detectable by either LISA or OMEGA.

\placefigure{snrplot}

The population of close binary systems expected to be found in globular clusters is an open question in the modeling of the dynamics of globular clusters (see \markcite{meylan} Meylan \& Heggie 1997; \markcite{McMillan} McMillan et al. 1997; and \markcite{hut} Hut et al. 1992 for reviews of the issue.) Observations of globular clusters have discovered binary populations in many individual clusters, such as NGC 2808 \markcite{ferraro} (Ferraro et al. 1997), NGC 6752 \markcite{rubenstein} (Rubenstein \& Bailyn 1997), and M30 \markcite{alcaino} (Alcaino et al. 1998). Due to the difficulties inherent in identifying binaries in such crowded systems, the populations discovered so far suffer significantly from selection effects. Several numerical models have been developed to determine expected binary populations, but the complexity of the issue requires that many approximations and assumptions be made. N-body simulations can currently handle systems of up to 32,000 bodies, but fall short of the $10^5$ to $10^6$ bodies needed to model full globular clusters \markcite{portegies1} (Portegies Zwart et al. 1997a). For a good summary of N-body simulations including stellar evolution, see \markcite{portegies3} Portegies Zwart et al. (1997c). Other approaches involve assuming simplified models of mass distribution, stellar evolution, or both (\markcite{kim} Kim et al. 1998; \markcite{vesperini} Vesperini \& Chernoff 1996; and \markcite{sigurdsson} Sigurdsson \& Phinney 1995). In many cases, these results serve to demonstrate the likelihood of a scenario producing specific types of binary systems rather than to generate a model population distribution in both stellar type and frequency. The numerical model of \markcite{davies} Davies (1995) incorporates a significant number of realistic assumptions and generates several model binary populations for a variety of globular cluster and stellar evolution models. Unfortunately, there is no frequency distribution given, but instead a variety of end products are listed for each model.

A crude model for the frequency distribution in the population of very short period binaries can be obtained from Davies' results by considering the results of two different runs in Table 10 of \markcite{davies} Davies(1995). In one run, the evolution of 1000 binaries are followed through 15 Gyr in a typical globular cluster. The outcomes of the evolution include four classes of binaries which merge due to gravitational radiation induced inspiral: neutron stars/white dwarf (NW), white dwarf/white dwarf with total mass above the Chandrasekhar mass ($WD_a^2$) and total mass below the Chandrasekhar mass ($WD_b^2$), and neutron star/neutron star ($NS^2$). In the second run, the exact same model is used, except that the evolution is followed until all the binaries end up in either merged or contact systems. The results for the four types of mergers are shown in Table~\ref{davies} \markcite{davies} (Davies 1995).

\placetable{davies}

At the end of 15 Gyr, one can assume that the population of short period binary systems consists of those which will merge later. The period distribution of this population can then be determined by assuming a constant merger rate until the entire population has merged. We assume that the population will have merged within an additional 15 Gyr, and so the merger rate per 1000 primordial binaries can be determined. The evolution of the period of a binary due to gravitational radiation losses is given by \markcite{hils} (Hils et al. 1990):
\begin{equation}\label{pdot}
\frac{dP}{dt} = -k_o P^{-5/3}
\end{equation}
where $k_o$ is given by:
\begin{equation}
k_o = \frac{96}{5} (2 \pi)^{8/3} \frac{G^{5/3}}{c^5} m_1 m_2 \left(m_1 + m_2\right)^{-1/3}.
\end{equation}
The period distribution is then related to the merger rate $\eta$ by:
\begin{equation}
\eta dt = \frac{dn}{dt} dt = \frac{dn}{dP}\frac{dP}{dt} dt
\end{equation}
and so,
\begin{equation}
\frac{dn}{dP} = \frac{\eta}{dP/dt} = \frac{\eta}{k_o} P^{5/3}.
\end{equation}

It then remains to determine the number of primordial binaries present in a globular cluster to estimate the current period distribution. Estimates of the primordial binary fraction range from 0 to 65\%, with 10\% considered a lower bound and 30\% considered reasonable (\markcite{yan1} Yan \& Mateo 1994; \markcite{yan2} Yan \& Reid 1996; \markcite{yan3} Yan \& Cohen 1996). For a large cluster such as 47 Tuc, the number of stars can be estimated by the mass of the cluster, so we assume a population of $10^6$ stars giving $3 \times 10^5$ primordial binaries. The number of close binary systems expected to merge due to inspiral is then 300 times the numbers given in Table~\ref{davies}. We can then find the expected period distribution based upon the merger rate. The number of binaries expected to exist with a period below 2000 s is then given by:
\begin{equation}
N = \int_{P_c}^{2000}\frac{\eta}{k_o} P^{5/3} dP
\end{equation}
where $P_c$ is given in Table~\ref{ddbtype} and the white dwarf is assumed to be CO. The results for a cluster of $10^6 M_{\sun}$ and for the globular cluster system as a whole with $M = 10^{7.5} M_{\sun}$ \markcite{sigurdsson} (Sigurdsson \& Phinney 1995) are presented in Table~\ref{numbers}. Thus, the globular cluster system may produce approximately 300 systems which can be detected with either LISA or OMEGA and whose orbital periods are low enough to be potentially identifiable with individual globular clusters.
\placetable{numbers}

\section{Directional Detection of Gravitational Radiation}
The orbital motion of both proposed detectors carries them about the sun with an orbital period of one year. This motion causes a periodic Doppler shift to appear in the frequency of any continuous source detected. From the details of this periodic shift, the angular position of the source can determined. The Doppler shift is given by:
\begin{equation}
f^{\prime}(t) = f_{gw}\sqrt{\frac{1 + {\bf v_e \cdot \hat{r}_s}/c}{1 - {\bf v_e \cdot \hat{r}_s}/c}}
\end{equation}
where ${\bf v_e}$ is the earth's velocity in the sun's rest frame, and ${\bf \hat{r}_s}$ is the unit vector joining the sun and the source. Because $v_e/c \sim 10^{-4}$, this can be written:
\begin{equation}
f^{\prime}(t;\theta_s,\phi_s) \simeq f_{gw} (1 + \frac{v_e}{c} \sin\theta_s \sin{(\omega t - \phi_s)})
\end{equation}
where $\omega = 2\pi yr^{-1}$, $\theta_s$ is the angle that ${\bf \hat{r}_s}$ makes with the normal to the earth's orbital plane, and $\phi_s$ is the angle that ${\bf \hat{r}_s}$ makes with the line joining the sun and the detector at $t = 0$. The process of determining the angular position of the source is then reduced to determining the parametrization of the detected signal:
\begin{equation}
h(t;A,f_gw,\theta_s,\phi_s,\psi) = A f^{\prime 2/3}\cos{(2\pi f^{\prime} t + \psi)}
\end{equation}
with $\psi$ the phase of the signal and $A$ the frequency independent part of the amplitude. In principle, the Doppler shift in the frequency can be detected if it is greater than the width of the frequency bins. The bin width is given by $\Delta f = 1/T_o$ where $T_o$ is the observations time and thus the minimum frequency is given by:
\begin{equation}
f_{gw} \geq \frac{c}{v_e T_o} = \frac{c}{\lambda_o}
\end{equation}
where the maximum wavelength is $\lambda_o = n \pi$ A.U. for an observation time of n years. This minimum is best interpreted as a lower bound for which the source may be identified with one half of the sky. Reasonable angular resolutions will require somewhat higher frequencies (see e.g. Cutler 1998).

The accuracy with which the parametrization of a signal can be determined has been discussed in a variety of papers (\markcite{finn}Finn 1992; \markcite{chernoff}Finn \& Chernoff 1993; \markcite{cutlerflanagan}Cutler \& Flanagan 1994; \markcite{schutz2}Schutz 1997b; \markcite{pjds}Peterseim et al. 1997 and \markcite{cutler}Cutler 1998). Although the precise description of the accuracy with which these parameters can be measured depends upon specific details of the detectors' response functions and orientation with respect to their orbit, general (although more approximate) results can be obtained. To first order, the accuracy of any measurement is inversely proportional to the signal to noise ratio ($\rho$). Thus the uncertainty in solid angle $\Delta\Omega = \sin{\theta_s} \Delta\theta_s \Delta\phi_s$ is proportional to $\rho^{-2}$. However, the region in the sky described by $\Delta\Omega$ will, in general, be elliptical due to the correlation between $\theta_s$ and $\phi_s$. Two recent papers have determined angular resolutions for LISA using semi-analytic techniques. Peterseim et al. (1997) determine a maximum uncertainty in $\Omega$ of $16 \mu sr$ for a plane wave of frequency $f_{gw} = 3 mHz$ and $\rho = 115$. \markcite{cutler}Cutler (1998) provides a number resolutions for specific frequencies and angular positions with $\rho = 10$. Comparing the results of Cutler (1998) with those of Peterseim et al. (1997) by scaling up the corresponding signal to noise ratio indicates that Cutler's results are roughly a factor of 5 larger.

Table~\ref{16gcs} lists the 16 closest globular clusters, which all lie within roughly 5 $kpc$. The angular separation between each cluster and any other cluster in the Milky Way globular cluster system is also listed and ranges from 14 to 270 $mrad$. In order to identify sources of gravitational radiation with specific globular clusters, the angular resolution of either LISA or OMEGA must be within this angular separation. The angular resolution for a given signal to noise ratio is approximated from the results of Cutler (1998)\markcite{cutler} by $\Delta\Omega \simeq \Delta\Omega_o \rho_o^2/\rho^2$ where $\Delta\Omega_o$ and $\rho_o$ are the values given by Cutler. The minimum signal to noise ratio needed to identify a source with a globular cluster within an angular separation of $\alpha$, is found by:
\begin{equation}
\rho = \frac{\rho_o}{\alpha} \sqrt{\Delta\Omega_o}
\end{equation}
Cutler uses $\rho_o = 10$, and finds $\Delta\Omega \simeq 0.04 sr$ for a gravitational wave with frequency $f_{gw} = 10^{-3} Hz$. For $\alpha = 20 mrad$, this requires $\rho \geq 100$. Since angular resolution improves with increasing frequency, the required signal to noise ratio decreases for higher frequencies, dropping to 15 for $f_{gw} = 10^{-2} Hz$. Since this signal to noise ratio is comparable to that shown in Figure~\ref{snrplot}, both neutron star-neutron star binaries and neutron star-white dwarf binaries will be readily identifiable with globular clusters out to $4 kpc$.

\placetable{16gcs}

At high frequencies, some binary systems will be losing enough energy in gravitational radiation so that its frequency will change (or chirp) measurably during the observation time. In this case, it is possible to determine the distance to the chirping binary. The rate at which the orbital period decreases is directly related to the absolute luminosity (in gravitational radiation) of the system. The observed signal strength is related to the apparent luminosity of the system. Thus, the distance to the binary is determined through comparing the apparent and absolute luminosities using (Schutz, 1997a)\markcite{schutz1}:
\begin{equation}\label{distance}
r = \frac{c}{2{\pi}^2} \frac{\dot{f}_{gw}}{f_{gw}^{3} h},
\end{equation}
where $\dot{f}_{gw}$ is the rate of change of the frequency. Note that this result does not require that the mass of the system be known. A lower bound can be placed upon the detectable value of $\dot{f}_{gw}$ by noting that the frequency must shift through one frequency bin during the observation time, $T_o$. Thus we have $\dot{f}_{gw} \geq T_o^{-2}$. Using Eq~\ref{strainamp} and Eq~\ref{pdot} we can find lower bounds on the frequency and signal to noise ratio for binaries whose chirp can be detected. Table~\ref{chirps} summarizes these bounds for the degenerate binaries given in Table~\ref{ddbtype}. As can be seen, the signal to noise ratio for any chirping binary in a nearby globular cluster will be above 100 with the exception of the He-He binaries.
\placetable{chirps}

In the frequency window of either OMEGA or LISA, any system in a globular cluster which chirps will be expected to have a very slowly varying frequency. Furthermore, the signal from a chirping binary will be strong enough to allow a reasonably confident identification with a specific globular cluster. Thus, the angular positions of the source ($\theta_s$ and $\phi_s$) can be considered to be known {\it a priori}. The expected signal from a chirping binary can then be written:
\begin{equation}
h(t;A,f_{gw},\dot{f}_{gw},\psi) = A (f_{gw}+\dot{f}_{gw} t)^{2/3}\cos{(2\pi f_{gw}t+\pi\dot{f}_{gw}t^2+\psi)}
\end{equation}
where positional information has been incorporated into $A$, and $\psi$ is an arbitrary phase. Once again, the process of determining the distance to the chirping binary is a matter of determining the parametrization of the signal. Although the details again require information on the specifics of the detectors' response functions, the uncertainty in the distance measurement will again be proportional to $1/\rho$. Previous work on determining the chirp frequency has focused primarily on the mergers of black hole binaries with the intended applicability being toward interpretation of LIGO data (e.g. Finn \& Chernoff 1993). Cutler (1998) has done some numerical estimations of distance measurements to supermassive black hole mergers which will inspiral in the frequency window of LISA. His results indicate an uncertainty in distance measurements ranging from 0.1\% to 30\% for signal to noise ratios much higher than we can expect from degenerate binaries in globular clusters. However, large uncertainties in distance are correlated with large uncertainties in angular position. In the case of globular clusters, the angular position is assumed and so there should be a corresponding improvement in the distance estimations. For the mergers of two $10^4 M_{\sun}$ black holes, Cutler finds signal to noise ratios ranging from 148 to 698 and uncertainties in the distance ranging from 0.8\% to 29\%. Discarding the highest and lowest uncertainties, we find $\Delta r/r \sim 7/\rho$. Using this, we find lower bounds on uncertainties in the distance measurements to globular cluster to be within the range of 1\% for NS-NS binaries to 10\% for He-He binaries. Clearly, there is promise in using chirping binaries to measure globular cluster distances.

\section{Observational Evidence of Binaries in Globular Clusters}
Observational evidence of binaries in globular clusters comes from a variety of surveys, each with its own limitations. Photometric surveys, which are restricted to studies of short period ($P \leq 5$ days), have discovered eclipsing binaries and cataclysmic variables. Radial velocity surveys have been applied to luminous giants and are therefore biased towards binary systems with periods greater than about 40 days. One can also search for a binary sequence in a color-magnitude diagram for a given globular cluster\markcite{ferraro} (Ferraro et al. 1997), although this applies only to main sequence binaries. Currently, the results point to a main sequence binary frequency in globular clusters to be around 30\% which is lower than the solar neighborhood frequency of about 65\%\markcite{meylan} (Meylan \& Heggie, 1997). However, some observations yield much lower binary frequencies for main sequence stars in specific globular clusters\markcite{alcaino} (Alcaino et al. 1998).

Binary systems are thought to be energy sources which delay the onset of core collapse during the evolution of a globular cluster. In interactions with other stars and binaries in the cluster hard binaries tend to become harder, releasing gravitational energy to heat the cluster. In a three-body interaction, the most common result is the ejection of the least massive star of the three, leaving the two more massive stars in a tighter orbit. The net result of these interactions is to increase the number of massive, hard binaries found in the core, and the ejection or destruction of softer binaries. This separation can be seen in the bimodal frequency in NGC 6752, which has an inner core frequency of around 25\% and an outer frequency of less than 16\%\markcite{rubenstein} (Rubenstein \& Bailyn, 1997). The binaries which remain in the core will have had a variety of evolutionary pathways available to form more exotic binary systems containing at least one collapsed object.

Globular clusters contain some of the oldest populations of stars in the galaxy, and so the mass function of non-degenerate objects extends only to about $0.8 M_{\sun}$. Consequently, one would expect there to be many dark, compact objects in a globular cluster population. These objects would only be detectable through gravitational effects unless they were interacting through accretion with a binary companion. If one were to apply a standard Salpeter initial mass function to a bright globular cluster (e.g. 47 Tuc), one would find approximately 5000 neutron star progenitors, 2000 black hole progenitors, and 300,000 white dwarf progenitors\markcite{McMillan} (McMillan et al. 1997). Recent estimates of the white dwarf mass fraction of NGC 6121 give a value of $0.15 \pm 0.10$ with a mean mass of the white dwarfs at $0.55 M_{\sun}$\markcite{hippel} (von Hippel 1998). It is quite probable that many of these progenitors evolved in binary systems or have entered binary systems through interaction and may make up a population of double degenerate systems which are very difficult to detect. This population should contain several members which will be bright enough in gravitational radiation to be seen by either LISA or OMEGA.

There have been several searches for exotic binaries in globular clusters. In searches using electromagnetic radiation, these exotic objects are most likely to be found as pulsars, low mass X-ray binaries (LMXBs) or cataclysmic variables (CVs). Kulkarni \& Anderson\markcite{kulkarni} (1996) list 33 known pulsars in globular clusters, 14 of which are known to be in binary systems. There are also at least 12 bright cluster LMXBs known\markcite{verbunt} (Verbunt et al. 1995) and the number of CV candidates is at 13\markcite{McMillan} (McMillan et al. 1997). These observations point to the existence of a population of exotic binaries which are predicted by evolutionary models of globular cluster dynamics\markcite{hut} (Hut et al. 1992). There are particular examples of extremely hard binaries in globular clusters. The shortest known binary system (4U 1820-30) is an X-ray source in NGC 6624. This system has an orbital period of 685 s and is thought to consist of a neutron star accreting matter from a degenerate helium dwarf of $0.07 M_{\sun}$\markcite{morgan} (Morgan et al. 1988). There is also an X-ray source "Star S" in NGC 6712 which has recently been determined to be a double degenerate system with an orbital period of 20.6 minutes\markcite{homer} (Homer et al. 1996). Thus, two of the three best studied optical counterparts to X-ray sources in globular clusters have been shown to be ultra-short period double degenerate systems\markcite{anderson} (Anderson et al. 1997).

\section{Discussion}
With the above evidence for a globular cluster population of binaries which are observable by LISA or OMEGA, it remains only to determine which clusters are the best candidates for observation. The cluster should be nearby so that the signal from binaries is strong and it should be large an populous so that there is a larger number of possible binaries. In addition, clusters which are significantly removed from the galactic disk and well separated from neighboring globular clusters will reduce the detection of possible field binaries which must be discriminated from the cluster binaries. Of the sixteen clusters listed in Table~\ref{16gcs}, five are good candidates. The two best are NGC 6752 and NGC 104 (47 Tuc)\markcite{harris1} (Harris 1998).

NGC 6752 is a post-core-collapse cluster only 3.9 kpc distant, with a large main-sequence binary frequency in the core. Consequently, it should also have a large population of cataclysmic variable, however the large stellar density of the core makes their identification difficult. There have been two candidate CVs identified with orbital periods of 5.1 and 3.7 hr\markcite{bailyn} (Bailyn et al. 1994). NGC 104 is a large globular cluster located at a distance of 4.3 kpc and known to have a large population of exotic binaries. Five of the eleven pulsars known to inhabit 47 Tuc have been identified as binaries and three of them are thought to have orbital periods under one day\markcite{kulkarni} (Kulkarni \& Anderson 1996). The only known dwarf nova in the core of a globular cluster lies in the core of 47 Tuc\markcite{paresce} (Paresce \& De Marchi 1994).

The angular position of NGC 6752, measured with respect to the normal to the plane of the ecliptic, is $\theta_s = 127\arcdeg$. The angular position within the plane of the ecliptic, $\phi_s$ is determined in part by the position of the detector at the beginning of the observation, which cannot be determined until the missions are actually flown. However, it is expected that either mission is expected to take data for more than one year. Therefore, the year-long data stream can be begun at a time when $\phi_s$ provides the most favorable angular resolution. Using Peterseim et al. (1997), a 3mHz source with $\rho = 115$ can be localized to within a $3 mrad \times 3 mrad$ patch in the sky. The nearest globular cluster to NGC 6752 is separated by $190 mrad$, and is $12.9 kpc$ distant. The corresponding angular position of 47 Tuc is $\theta_s = 152\arcdeg$, and so Peterseim et al. gives $\Delta\theta_s = 2.5 mrad$ and $\Delta\phi_s = 5 mrad$ for a 3mHz source with $\rho = 115$. 47 Tuc is separated by $63 mrad$ from its nearest neighbor, which is $8.3 kpc$ distant.

If the dense core of either of these clusters hold a population of unidentified short period degenerate binaries, they should be easily detectable during the course of a year-long observation in gravitational radiation. The strategy to follow in order to detect the binary population of a globular cluster would be to search the data stream from the detector using a template for which the angular position of the source is {\it a priori} identified with the globular cluster. If a number of sources appear and they have signal-to-noise ratios above 100, then these sources can be reasonably assigned to the globular cluster. If any of these sources have a frequency higher than the chirp frequency limits given in Table~\ref{chirps}, then a template incorporating a chirp can be applied to the data stream in search of a detectable chirp. If a chirp is detected, then the distance to the globular cluster can be determined using Eq~\ref{distance}. The signal-to-noise ratios given in Table~\ref{chirps} indicates that any detectable chirping binary should provide distance measurements to better than 10\% accuracy.

\section{Conclusion}
Space-based gravitational radiation detectors are expected to provide a wealth of information concerning the basic nature of gravity. By observing radiation from coalescing black holes, neutron stars, and supermassive black holes, they should help map out the space time geometry around black holes and provide high precision tests of general relativity. In addition to adding to the body of knowledge concerning gravitation theory, these detectors can also allow for the study of more prosaic objects such as white dwarf and neutron star binaries. These objects, in turn, can be used to study the astronomical objects which are expected to contain them. By observing short period binary populations in the dense cores where conventional electromagnetic observations are hampered by crowding, better knowledge of the nature of the core binary populations of globular clusters can be obtained. This will provide information about the evolution and structure of such objects. It may also be possible to determine the distance to several globular clusters, providing a means to obtain independent distance measurements to hundreds of RR-Lyrae stars and other conventional astronomical distance indicators. Because globular clusters are expected to harbor many binary systems, the tedious task of all-sky searches for sources of gravitational radiation can be avoided by concentrating on the locations where they are most likely to be. Globular clusters may be the first area where basic astronomy observations can be done in gravitational radiation.

\acknowledgments

This work was supported in part by NASA EPSCoR grant NCCW-0058 and Montana EPSCoR grant NCC5-240.

\clearpage
\begin{figure}
\figcaption{Signal strengths of binaries listed in Table~\ref{ddbtype} from a globular cluster at 4 kpc distance. Note that the NS-He and CO-CO lines are nearly coincidental, with the NS-He signal just slightly stronger. The instrument noise levels are taken from Hiscock (1998) and are for a one year observation.\label{snrplot}}
\end{figure}

\clearpage
\begin{deluxetable}{lrrr}
\tablecaption{Double Degenerate Binary Properties. \label{ddbtype}}
\tablewidth{0pt}
\tablehead{
\colhead{Binary Type} & \colhead{$m_1 (M_{\sun})$} & \colhead{$m_2 (M_{\sun})$} & \colhead{$P_c (s)$}
}

\startdata
NS-NS & 1.4 & 1.4 & $10^{-3}$\nl
NS-CO & 1.4 & 0.6 & 80\nl
NS-He & 1.4 & 0.3 & 125\nl
CO-CO & 0.6 & 0.6 & 100\nl
CO-He & 0.6 & 0.3 & 160\nl
He-He & 0.3 & 0.3 & 200\nl
\enddata
\end{deluxetable}

\clearpage
\begin{deluxetable}{lrrrr}
\tablecaption{Inspiral Merger Rates. \label{davies}}
\tablewidth{0pt}
\tablehead{
\colhead{$T_{evol}$} & \colhead{NW} & \colhead{$WD_a^2$} & \colhead{$WD-b^2$} & \colhead{$NS^2$}
}

\startdata
15 Gyr & 10 & 11 & 0 & 1\nl
$\infty$ & 74 & 57 & 0 & 18\nl
\enddata
\end{deluxetable}

\clearpage
\begin{deluxetable}{lrrr}
\tablecaption{Short Period Binaries. \label{numbers}}
\tablewidth{0pt}
\tablehead{
\colhead{} & \colhead{NW} & \colhead{$WD_a^2$} & \colhead{$NS^2$}
}

\startdata
Cluster & 4.0 & 5.6 & 0.5\nl
GC System & 125.2 & 176.5 & 16.0\nl
\enddata
\end{deluxetable}

\clearpage
\begin{deluxetable}{lrlrr}
\footnotesize
\tablecaption{Closest Globular Clusters\tablenotemark{a}. \label{16gcs}}
\tablewidth{0pt}
\tablehead{
\colhead{Cluster ID} & \colhead{Dist ($kpc$)} & \colhead{Nearest Neighbor} & \colhead{$\triangle \varphi$ ($mrad$)} & \colhead{Neighbor Dist. ($kpc$)}
}
\startdata
NGC 6121 & 2.1 & NGC 6144 & 17 & 10.1\nl
NGC 6397 & 2.2 & NGC 6584 & 91 & 12.9\nl
NGC 6544 & 2.5 & NGC 6553 & 35 & 4.7\nl
NGC 6656 & 3.2 & NGC 6642 & 15 & 7.7\nl
NGC 6540 & 3.5 & Djorg 2 & 15 & 13.8\nl
NGC 6366 & 3.6 & IC 1257 & 36 & 24.4\nl
NGC 6838 & 3.8 & Pal 10 & 150 & 5.8\nl
NGC 6752 & 3.9 & NGC 6584 & 190 & 12.9\nl
E 3 & 4.2 & NGC 2808 & 220 & 9.3\nl
NGC 6254 & 4.3 & NGC 6218 & 55 & 4.8\nl
NGC 104 & 4.3 & NGC 362 & 63 & 8.3\nl
NGC 6553 & 4.7 & Terzan 10 & 14 & 8.3\nl
NGC 6218 & 4.8 & Pal 15 & 52 & 43.6\nl
NGC 4372 & 5.0 & NGC 4833 & 56 & 5.8\nl
NGC 5139 & 5.1 & NGC 5286 & 82 & 10.7\nl
NGC 3201 & 5.1 & Pyxis & 270 & 38.5\nl
\enddata
\tablenotetext{a}{Data taken from \markcite{harris} Harris (1996)}
\end{deluxetable}

\clearpage
\begin{deluxetable}{lcrr}
\tablecaption{Chirp Frequency Limits. \label{chirps}}
\tablewidth{0pt}
\tablehead{
\colhead{Binary Type} & \colhead{$f_c (mHz)$} & \colhead{$\rho$} & \colhead{$P (s)$}
}

\startdata
NS-NS & 3.7 & 523 & 540\nl
NS-CO & 4.6 & 286 & 434\nl
NS-He & 5.4 & 169 & 370\nl
CO-CO & 5.5 & 165 & 363\nl
CO-He & 6.4 & 101 & 312\nl
He-He & 7.5 & 64 & 267\nl
\enddata
\end{deluxetable}

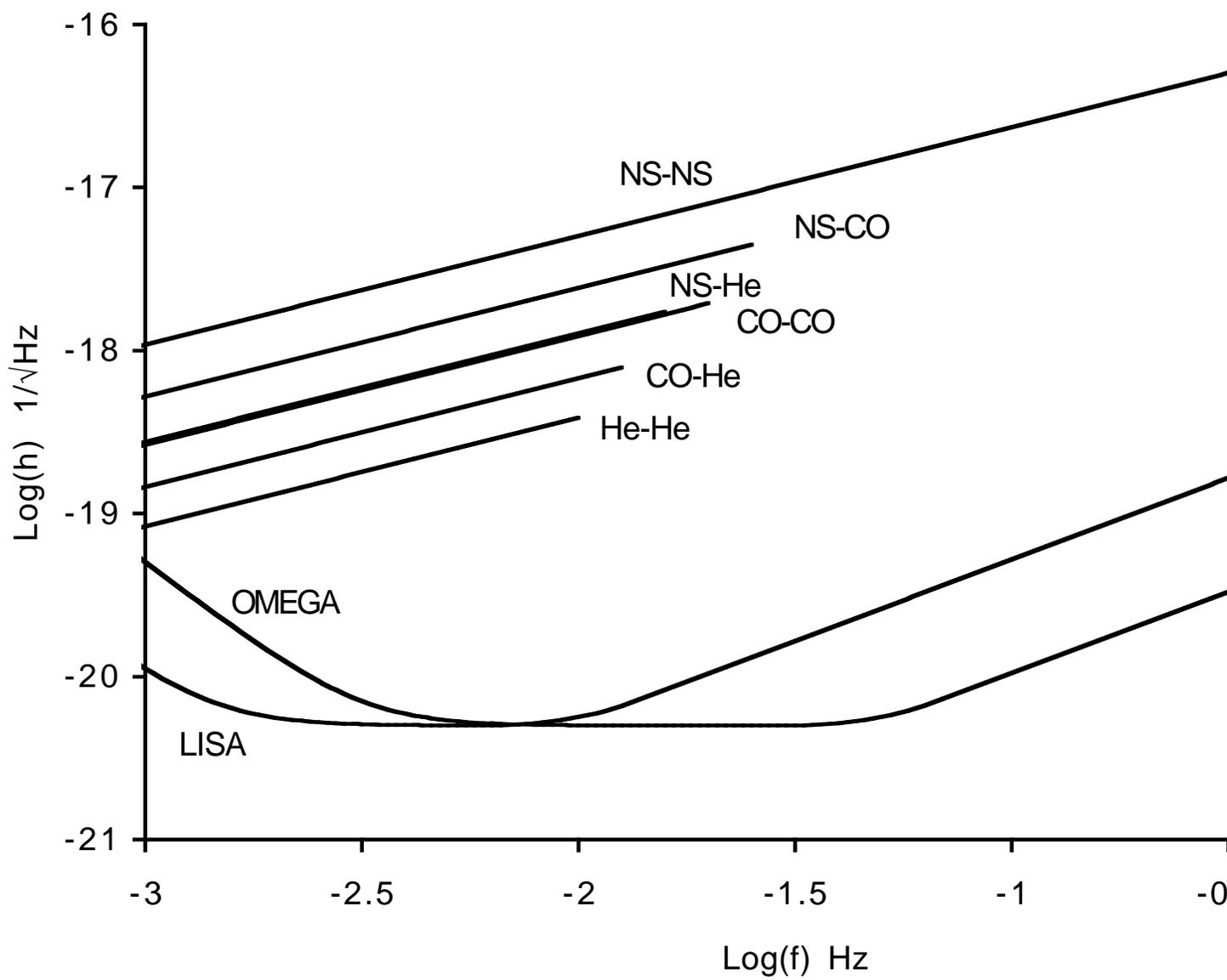
\end{document}